\documentclass[aps,pra,twocolumn,showpacs,superscriptaddress,bibliography]{revtex4-1}
\usepackage{graphicx}    % needed for figures
\usepackage{dcolumn}    % needed for some tables
\usepackage{bm}             % for math
\usepackage{amssymb}   % for math
\usepackage{amsmath}    % for multiline equation
\usepackage{subfigure}    % for subfigure
\usepackage{braket}
\usepackage{mathrsfs}
\usepackage{multirow}
\usepackage[export]{adjustbox}
\usepackage[toc,page]{appendix}
\begin{document}

\title{Dressing trapped ions with integrated wires}

\author{R. T. Sutherland}
\email{tyler.sutherland@oxionics.com}
\address{Oxford Ionics Limited, Unit 1, Oxford Technology Park, Technology Dr, Kidlington OX5 1GN, United Kingdom}
\address{Quantinuum, 303 S Technology Ct, Broomfield, CO 80021, USA}

\date{\today}

\begin{abstract}
We discuss dressing trapped ions with the near field of a trap integrated wire. Ramping a dressing field on/off adiabatically before/after an operation changes its effective Hamiltonian. The amplitude and detuning of the dressing field act as tunable degrees of freedom we can use to `customize' the properties of any operation. We propose three use cases for this general tool. First, we can generate `artificial' clock states, where we eliminate the (assumed to be small) linear sensitivity of a qubit. Second, we can break the degeneracies that often complicate shelving at low quantization fields\textemdash allowing us to implement operations with linearly polarized microwaves that would, otherwise, require circular polarization. Finally, we can implement laser-free single qubit gates on a set of `target' ions using fields that are separated from the rest of the computer in frequency space.
\end{abstract}
\maketitle

\section{Introduction}
\begin{figure}[b]
\includegraphics[width=0.5\textwidth]{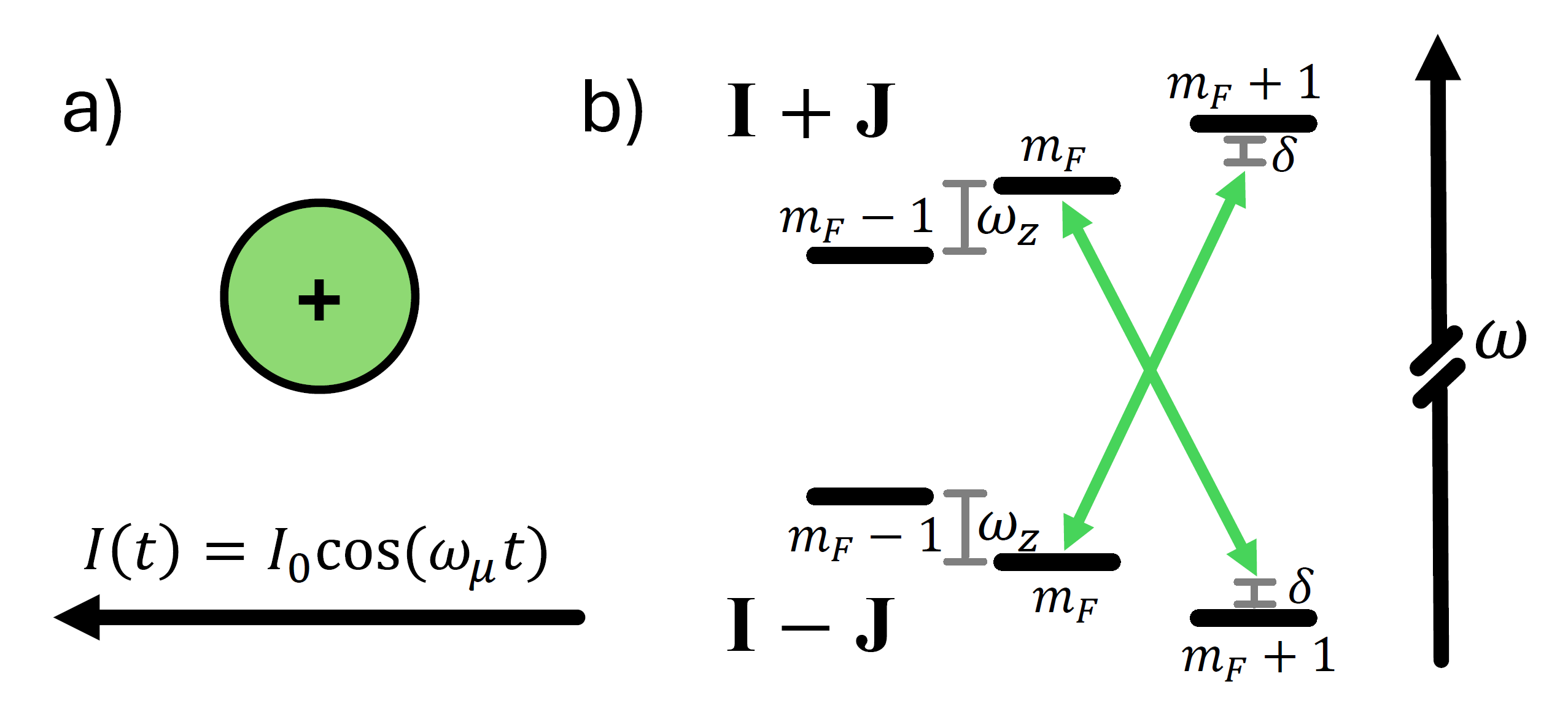}
\centering
\caption{(a) We shuttle the target ion near an integrated wire, then adiabatically ramp on/off a $\propto \omega_{\mu}$ AC current, using its near-field to dress the atom before/after an operation. (b) Level diagram of a dressed hyperfine manifold. Two pairs of states are simultaneously dressed due to the near-degeneracy of the two transitions at low quantization fields.}
\label{fig:fig_1}
\end{figure}

Trapped ions are a promising qubit because of many of their demonstrated properties: high fidelity gates, long memory times, and the ability to simplify circuits with all-to-all connectivity \cite{wineland_1998, ladd_2010, harty_2014, ballance_2016,gaebler_2016, srinivas_2021}. These demonstrations, however, typically occur in stand-alone laboratory experiments, and combining them into one device is its own challenge \cite{pino_2021,malinowski_2023,moses_2023}. Any qubit scheme will, ultimately, have benefits and drawbacks. For example, using an intermediate field clock state \cite{harty_2014,harty_2016} makes shelving straightforward, since the degeneracies that complicate schemes at lower quantization fields are broken in that regime. On the other hand, fractional fluctuations in the quantization field\textemdash over the entire computer\textemdash map onto larger fluctuations in the magnetic fields experienced by the ions. In this work, we focus on addressing some of the drawbacks associated with qubits in `low' $B_{q}\lesssim~10~\text{Gauss}$ quantization fields. For example, consider the $S_{1/2}$ ground state hyperfine manifold of a nuclear spin-$I$ ion with a $J=1/2$ spin valence electron. The pair of $\ket{F^{\pm}\equiv I\pm J,m_{F}=0}$ states is a popular choice of qubit \cite{pino_2021,moses_2023}, which, unfortunately, is only a true `clock' qubit (zero linear sensitivity to magnetic fields) at exactly zero quantization field $B_{q}=0$. This means the system will have longer memory times for smaller values of $B_{q}$. If we want to perform logical operations, however, the Zeeman splittings approach zero when $B_{q}\rightarrow 0$, increasing the risk of state leakage outside the qubit manifold and limiting gate speed. Therefore, if the qubits experience (approximately) the same magnetic environment over the entire computer, state leakage and memory errors will act as competing bottlenecks\textemdash a problem absent from any isolated demonstration. Another challenge when using $B_{q}\lesssim~10~\text{Gauss}$ quantization fields is that many relevant shelving transitions are nearly degenerate in this regime. For example, the transition $\ket{F^{+},m_{F}}\leftrightarrow\ket{F^{-},m_{F}\pm 1}$ is nearly-degenerate with $\ket{F^{-},m_{F}}\leftrightarrow\ket{F^{+},m_{F}\pm 1}$ (see Fig.~\ref{fig:fig_1}b), and their Rabi frequencies typically have irrational ratios. Theoretically, we could select these transitions using circularly polarized light, but this has not been demonstrated with microwaves and would be challenging to do at high-fidelities with lasers. Finally, another well-known example of this problem is laser-free one-qubit gates. While the authors of Ref.~\cite{harty_2014} showed it is possible to reach $\lesssim 10^{-6}$ infidelities for microwave one-qubit gates, their scheme (Rabi flopping) requires driving the targeted qubit with a magnetic field oscillating at the qubit transition frequency $\omega_{0}$. Since the wavelengths of microwaves are too big to focus the fields on specific targets, this scheme would result in an unacceptable amount of crosstalk in any large system. Because of this, scientists have proposed several techniques to fix the issue via active field cancellation \cite{aude_2014,aude_2017,leu_2023} or by locally separating the target qubit in frequency space \cite{wang_2009, johanning_2009, warring_2013_prl, weidt_2016, srinivas_2021,sutherland_2023}, none of which have reproduced similar fidelities. \\

\noindent Atoms are inherently identical when placed in identical environments. Considering the quantum charge-coupled device architecture (QCCD), however, where ions are shuttled to/from designated locations on the computer, it is not necessarily the case that every ion experiences the same environment. Trap integrated wires can non-trivially alter the magnetic fields experienced by a set of `target' ions $\sim 10$'s of $\mu\text{m}$ away from the wire \cite{ospelkaus_2008,ospelkaus_2011,warring_2013_prl,harty_2014,harty_2016,srinivas_2018, srinivas_2021,malinowski_2023,sutherland_2024}, while only perturbing `idle' atoms $\sim 100~\mu\text{m}$'s away from the wire. Reference~\cite{sutherland_2024} explored this for quasi-static fields, discussing the benefits of temporarily altering the size and direction of a set of target ions. Here, in the same vein, we discuss the benefits of locally ramping on a $\sim\mathrm{GHz}$ microwave dressing field (tuned near the hyperfine splitting and linearly polarized in the xy-plane), before/after a set of target ions interacts with a second control field. First, we show how to tune this dressing field to create an `artificial' clock state, rendering the target ions insensitive to magnetic field noise. Second, we show that, when combined with an additional $\sim \text{MHz}$ $z$-polarized magnetic field, dressing unlocks a straightforward way to perform shelving operations that would be difficult otherwise. Finally, we discuss how dressing enables laser-free single-qubit gates with fields that are frequency separated from the rest of the computer, suppressing crosstalk.

\section{Theory}

For this work, we consider a single `target' trapped ion qubit $10$'s of $\mu\text{m}$ (roughly an ion height) away from an integrated wire. We restrict the discussion to the ion's ground-state $S_{1/2}$ hyperfine manifold, i.e. the coupling between a non-zero spin-$I$ nucleus, and the $J=1/2$ spin of a valence electron, noting the results generalize to other other hyperfine manifolds. The magnetic coupling of the electron with the nucleus generates two sets of hyperfine sublevels with angular momentum $F^{\pm}\equiv I\pm J$, each having $2F^{\pm}+1$ hyperfine sublevels $m_{F}$ which we label $\ket{F^{\pm},m_{F}}$. We assume that $\hat{H}_{0}$ includes a permanent magnetic quantization field $\vec{B}_{q}$, defining the $z$-axis. This gives a Hamiltonian: 
\begin{eqnarray}\label{eq:h0}
    \hat{H}_{0}=\frac{\hbar A}{2}\vec{I}\cdot\vec{J} + \mu_{\mathrm{B}}B_{q}(g_{J}\hat{J}_{\mathrm{z}}+ g_{I}\hat{I}_{\mathrm{z}}),
\end{eqnarray}
where $\vec{L}\equiv (\hat{L}_{x},\hat{L}_{y},\hat{L}_{z})$ is an angular momentum vector operator $\vec{L}\in \{\vec{I},\vec{J}\}$, $g_{J(I)}$ is the $g$-factor of the electron(nucleus), $\mu_{\text{B}}$ is the Bohr magneton, and $A$ is the hyperfine constant. The Hamiltonian is not diagonal in this basis, but selection rules tell us the off-diagonal elements only couple states with the same $m_{F}$, which are separated by the hyperfine splitting $\sim \mathrm{GHz}$ \cite{edmonds_1996,langer_2006}. For simplicity, we assume the permanent quantization field is $\lesssim 10$ Gauss, making the effective coupling of these states $\sim\mathrm{MHz}$ and giving a small quadratic shift; we neglect these terms unless stated otherwise (see Sec.~\ref{sec:sensitivities}). \\

\subsection*{Dressed Hamiltonians}\label{sec:dressing}

We are interested in the behavior of a `target' ion that is significantly closer to one or more integrated wires than the rest of the ions in the computer. Let the Hamiltonian associated with the dressing field be $f(t)\hat{H}_{D}$, where $f(t)$ is an envelope function that smoothly increases(decreases) from 0(1) to 1(0) over a time $\tau$ before(after) an operation. After we ramp on $\hat{H}_{D}$, we apply a second `target' Hamiltonian $\hat{H}_{{t}}$, whose properties will change because of $\hat{H}_{\text{D}}$. This makes the full system Hamiltonian:
\begin{eqnarray}
    \hat{H}=\hat{H}_{0}+f(t)\hat{H}_{D}+g(t)\hat{H}_{t},
\end{eqnarray}
where $g(t)=0$ when $f(t)\neq 1$ and $g(t)=1$ when $f(t)=1$. We can transform the system into the eigenbasis of  $\hat{H}_{D}$ using the unitary operator $\hat{U}_{D}$:  
\begin{eqnarray}\label{eq:interaction_ham}
    \hat{H}_{I} &=& \hat{U}_{D}^{\dagger}\hat{H}\hat{U}_{D} + i\hbar\dot{\hat{U}}_{D}^{\dagger}\hat{U}_{D} \nonumber \\
    & \simeq & \hat{H}_{I,0} + g(t)\hat{H}_{I,t},
\end{eqnarray}
where, in the second line, we assumed the value of $\tau$ is large enough to negate the contributions of diabatic processes, dropping the $\propto i\hbar\dot{\hat{U}}^{\dagger}_{D}\hat{U}_{D}$ term. Additionally, we can use the fact that that $f(0)=f(t_{f})=1$ to conclude that $\hat{U}_{D}(0)=\hat{U}_{D}(t_{f})=\hat{I}$, meaning the time propagator resulting from Eq.~(\ref{eq:interaction_ham}) converges to the time propagator of the full system \cite{sutherland_2019}. \\

\subsection*{Microwave dressing in hyperfine manifolds}\label{sec:hyperfine_dressing}

We consider the low quantization field limit, by which we mean the limit where quadratic Zeeman shifts that arise due to $\vec{B}_{0}$ coupling the hyperfine manifolds is small relative to linear Zeeman shifts from the quantization field, i.e. $B_{q}\lesssim 10~$Gauss. Thus, unless specified otherwise (see Sec.~\ref{sec:sensitivities}), we take the frequency splitting of every $\ket{F^{+},m_{F}}$ to be $\omega_{+,m_{F}}\simeq m_{F}\omega_{z}$, and every $\ket{F^{-},m_{F}}$ to be $\omega_{-,m_{F}}\simeq -m_{F}\omega_{z}$, where $\omega_{z}$ is the Zeeman splitting of the system. It follows that the transition $\ket{F^{+},m_{F}}\rightarrow\ket{F^{-},m_{F}\pm 1}$ has a frequency $\omega_{0}\pm\omega_{z}$, which is, up to the quadratic shift, the same frequency as the transition $\ket{F^{-},m_{F}}\rightarrow\ket{F^{+},m_{F}\pm 1}$ (see Fig.~\ref{fig:fig_1}b); here, $\omega_{0}$ is frequency difference between the hyperfine manifolds. Because of this, if we apply an $x$-polarized magnetic field that oscillates near the hyperfine splitting:
\begin{eqnarray}\label{eq:h_mu}
    \hat{H}_{D} = \mu_{\text{B}}B_{q}g_{J}\hat{J}_{x}\cos([\omega_{0}+(2m_{F}\pm 1)\omega_{z}-\delta]t),
\end{eqnarray}
we cannot resonantly drive $\ket{F^{+},m_{F}}\leftrightarrow\ket{F^{-},m_{F}\pm 1}$ without also resonantly driving $\ket{F^{-},m_{F}}\leftrightarrow\ket{F^{+},m_{F}\pm 1}$. In general, the value of $\bra{F^{+},m_{F}}\hat{J}_{x}\ket{F^{-},m_{F}\pm 1}\neq  \bra{F^{-},m_{F}}\hat{J}_{x}\ket{F^{+},m_{F}\pm 1}$, and is often different by an irrational factor. This makes difficult to select a transition with resonant fields alone, even with composite pulses. \\

\noindent To illustrate the process, we make the rotating wave approximation with respect to any term oscillating near $\omega_{z}$ and $\omega_{0}$, keeping only the two $\sim \delta$ dressed transitions after transforming into the rotating frame with respect to $\hat{H}_{0}$. For a system initialized to $\ket{\psi(0)}=c_{0}\ket{F^{-},m_{F}} + c_{1}\ket{F^{+},m_{F}}$, the relevant Hamiltonian for the system is: 
\begin{eqnarray}\label{eq:micro_ham_approx}
    \hat{H}_{D} &= & \hat{H}_{a} + \hat{H}_{b} \\
    &=& \frac{\hbar}{2}\Big(\Omega_{a}\ket{F^{+},m_{F}\pm 1}\bra{F^{-},m_{F}}e^{\pm i\delta t} \nonumber \\
    &&+\Omega_{b}\ket{F^{+},m_{F}}\bra{F^{-},m_{F}\pm 1}e^{\pm i \delta t} + c.c., \Big). \nonumber 
\end{eqnarray}
In Eq.~(\ref{eq:micro_ham_approx}), there are two, uncoupled, two state subspaces $a\equiv \{\ket{F^{-},m_{F}},\ket{F^{+},m_{F}\pm 1}\}$ and $b\equiv \{\ket{F^{-},m_{F}\pm 1}\},\ket{F^{+},m_{F}}$; labeling these states $\{\ket{0},\ket{1}\}$, respectively, we can simplify each Hamiltonian by rewriting it in terms of its own Pauli operators. Applying the transformations:
\begin{eqnarray}
    \hat{U}_{\delta,\alpha}=\exp\Big(i\delta t \hat{\sigma}_{z,\alpha} \Big),
\end{eqnarray}
where $\alpha \in \{a,b\}$, we can then rewrite the each subspace in a time-independent frame-of-reference:
\begin{eqnarray}
    \hat{H}_{\alpha} = \frac{\hbar\Delta_{\alpha}}{2}\Big(\hat{\sigma}_{x,\alpha}\sin[\theta_{}]+\hat{\sigma}_{z,\alpha}\cos[\theta_{\alpha}] \Big),
\end{eqnarray}
where $\Delta_{\alpha}\equiv \sqrt{\Omega_{\alpha}^{2}+\delta^{2}}$ and $\tan(\theta_{\alpha})\equiv \Omega_{\alpha}/\delta$, each of which can be diagonalized with:
\begin{eqnarray}
    \hat{U}_{\theta,\alpha}=\exp\Big(-\frac{i\theta_{a}}{2}\hat{\sigma}_{y,\alpha} \Big),
\end{eqnarray}
after which we apply $\hat{U}_{\delta,\alpha}^{\dagger}$ to make it easier to account for that transformation, making the total transformation for each subspace:
\begin{eqnarray}\label{eq:diag_u}
    \hat{U}_{D,\alpha} =\hat{U}_{\delta,\alpha}^{\dagger}\hat{U}_{\theta,\alpha}\hat{U}_{\delta,\alpha}
\end{eqnarray}
and the full Hamiltonian:
\begin{eqnarray}\label{eq:dressed_sigmaz}
    \hat{H}_{I,D} = \frac{\hbar[\Delta_{a}-\delta]}{2}\hat{\sigma}_{z,a} + \frac{\hbar[\Delta_{b}-\delta]}{2}\hat{\sigma}_{z,b},
\end{eqnarray}
making $\hat{H}_{I,D}$ the total shift due to the dressing field.

\section{Use Cases}

\subsection{Clock states}\label{sec:sensitivities}

It has been shown that, in certain cases, dressing fields can render atomic states that are magnetically sensitive, insensitive \cite{timoney_2011,sark_2014}. The scheme, though, generates qubit splittings proportional to the dressing fields' Rabi frequencies, causing a strong dependence on control noise. In this section, we discuss a different technique that uses a dressing field to eliminate a qubit's linear sensitivity\textemdash when that sensitivity is already small. The primary benefit is that it can operate at AC shifts $\lesssim 1~\text{kHz}$. Since the associated shifts are smaller, fractional uncertainties in their size lead to smaller errors. \\

\begin{figure}[b]
\includegraphics[width=0.5\textwidth]{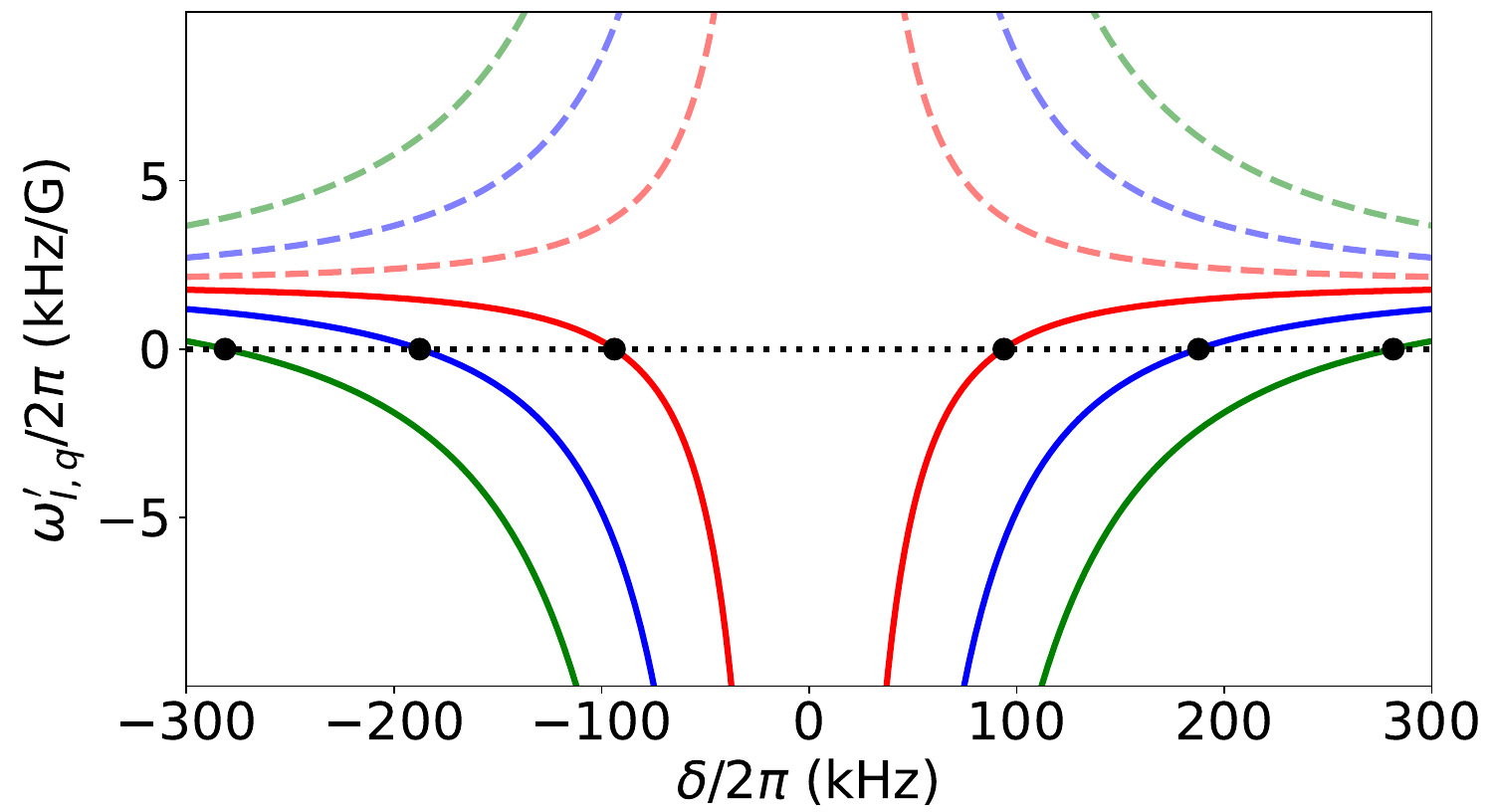}
\centering
\caption{Magnetic sensitivity $\omega_{I,q}^{\prime}$ of the $\{\ket{2,0},\ket{1,0} \}$ qubit versus dressing field detuning $\delta$ when we use a $10~\text{mG}$ (red center), $20~\text{mG}$ (blue second from center), and $30~\text{mG}$ (green furthest from center) microwave dressing field. For the solid curves, we dress the qubit with the $m_{F}=1$ states and for the dashed, we dress the qubit with $m_{F}=-1$. For the former, there is a value of $\delta$ where the linear magnetic sensitivity of the qubit vanishes, i.e. $\omega_{I,q}^{\prime}=0$. For each curve, we indicate the value of $\delta$ that Eq.~(\ref{eq:analytic_delta_clock}) predicts this will occur with a black dot.}
\label{fig:b_sense}
\end{figure}

\noindent We are often inclined to use qubits that have small, non-zero frequency sensitivities to magnetic fields. For example, in Refs.~\cite{pino_2021,moses_2023} the authors use states $\{\ket{0,0},\ket{1,0}\}$ of the $S_{1/2}$ manifold of $^{171}\text{Yb}^{+}$, which is a `clock' qubit only at $0~$Gauss. Since the ions need a quantization field to define the manifold and avoid leakage, we must apply a small ($\sim ~$Gauss) quantization field. This introduces a small ($\sim~\text{kHz}/\text{Gauss}$) linear sensitivity to the qubit, which significantly decreases memory times. We represent this with:
\begin{eqnarray}
    \hat{H}_{t,e} &=& \frac{\hbar\omega_{q}^{\prime}}{2}B_{e,z}\Big(\ket{F^{+},0}\bra{F^{+},0}-\ket{F^{-},0}\bra{F^{-},0} \Big), \nonumber \\
\end{eqnarray}
where $B_{e,z}$ is any non-accounted for magnetic field along $z$, and $\omega_{q}^{\prime}\equiv \partial \omega_{q}/\partial B_{z}$ is the linear sensitivity of the qubit. We can now use Eq.~(\ref{eq:diag_u}) to transform into `dressed' frame. Doing this and projecting onto the qubit manifold $\{\ket{F^{+},0},\ket{F^{-},0} \}$ gives:
\begin{eqnarray}
    \tilde{H}_{I,e}\!\!\! &=&\!\!\! \frac{\hbar\omega_{q}^{\prime}}{2}\!B_{z}\!\Big(\!\!\cos^{2}\!\!\Big[\frac{\theta_{b}}{2}\Big]\!\!\ket{F^{+}\!\!\!,0}\!\bra{F^{+}\!\!\!,0}\!-\!\cos^{2}\!\!\Big[\frac{\theta_{a}}{2}\Big]\!\!\ket{F^{-}\!\!\!,0}\!\bra{F^{-}\!\!\!,0}\! \Big). \nonumber \\
\end{eqnarray}
We can project $\hat{H}_{I,D}$ onto the qubit subspace as well:
\begin{eqnarray}
    \hat{H}_{I,D}&=& \frac{\hbar}{2}\Big(\Delta_{b}\ket{F^{+},0}\bra{F^{+},0}-\Delta_{a}\ket{F^{-},0}\bra{F^{-},0} \Big), \nonumber \\
\end{eqnarray}
which makes the total Hamiltonian:
\begin{eqnarray}
    \hat{H}_{I,t}&=&\hat{H}_{I,D}+\hat{H}_{I,e} \nonumber \\
    &=& \frac{\hbar}{2}\Big(\Big\{\omega_{q}^{\prime}B_{I,e}\cos^{2}\Big(\frac{\theta_{b}}{2}\Big)+[\Delta_{b}-\delta]\Big\}\ket{F^{+},0}\bra{F^{+},0} - \nonumber \\
    && \Big\{\omega_{q}^{\prime}B_{I,e}\cos^{2}\Big(\frac{\theta_{a}}{2}\Big)+(\Delta_{a}-\delta)\Big\}\ket{F^{-},0}\bra{F^{-},0}\Big) \nonumber \\
    &=& \frac{\hbar\omega_{q,I}}{2}\Big(\ket{F^{+},0}\bra{F^{+},0} - \ket{F^{-},0}\bra{F^{-},0}\Big) \nonumber \\
\end{eqnarray}
where, in the second line, we dropped a $\propto \hat{I}$ term and introduced:
\begin{eqnarray}\label{eq:dressed_mag_shift}
    \omega_{I,q}\!\equiv\! \frac{1}{2}\!\Big(\!\Big\{\omega_{q}^{\prime}B_{z}\!\cos^{2}\!\!\Big[\frac{\theta_{b}}{2}\Big]\!\!+\!\![\Delta_{b}-\delta]\Big\}\!\!  +\!\!  \Big\{\omega_{q}^{\prime}B_{z}\!\cos^{2}\!\!\Big[\frac{\theta_{a}}{2}\Big]\!\!+\!\![\Delta_{a}-\delta]\Big\} \Big), \nonumber \\
\end{eqnarray}
as the total qubit splitting. This makes the magnetic sensitivity:
\begin{eqnarray}\label{eq:dressed_mag_sense}
\omega_{I,q}^{\prime}\!\!\!\! &=&\!\!\!\frac{1}{2}\Big(\!\omega_{q}^{\prime}\!\Big\{\!\cos^{2}\!\!\Big[\frac{\theta_{b}}{2} \Big]\!\!+\!\cos^{2}\!\!\Big[\frac{\theta_{a}}{2} \Big]\! \Big\}\!\pm \!\omega_{z}^{\prime}\Big[\frac{\delta}{\Delta_{a}}\!+\!\frac{\delta}{\Delta_{b}}\!\mp\!2\Big]\Big), \nonumber \\ 
\end{eqnarray}
where we have allowed $\delta$ to be positive (red detuned) or negative (blue detuned). Here, we have used the fact that $\partial\delta/\partial B_{z} = \pm\omega_{z}^{\prime}$, where $\omega_{z}^{\prime}$ is magnetic sensitivity of $\ket{F^{+},\pm 1}$. Importantly, $\omega_{z}^{\prime}$ is positive when we dressing field couples the qubit to $\{\ket{F^{+}, 1},\ket{F^{-}, 1}\}$ and negative when it couples the qubit to $\{\ket{F^{+},-1},\ket{F^{-},-1}\}$. To render the system magnetically insensitive, we must ensure $\omega_{I,q}^{\prime}\simeq 0$, meaning we must tune the second term in Eq.~(\ref{eq:dressed_mag_sense}) to cancel the first. Since $\omega^{\prime}_{q}$ is always positive, this means Eq.~(\ref{eq:dressed_mag_sense}) has roots for both blue and red detuned dressing fields when coupling to $\{\ket{F^{+},1},\ket{F^{-},1}\}$ and \textit{none} when coupling to $\{\ket{F^{+},-1},\ket{F^{-},-1}\}$. There is an intuitive explanation for this. The quantization field quadratically pushes the qubit apart in frequency, making $\omega_{q}^{\prime}$ positive. At the same time, red(blue) detuned dressing fields also push the qubit apart(together) in frequency. When coupling to $\ket{F^{+},1},\ket{F^{-},1}$, this positive(negative) shift weakens(strengthens) with $B_{z}$, creating an additional negative dependence on $B_{z}$; we can tune this dependence to cancel $\omega_{q}^{\prime}$. When coupling to $\ket{F^{+},-1},\ket{F^{-},-1}$ on the other hand, red(blue) detuned light becomes less(more) detuned with $B_{z}$, creating the opposite effect, and so we cannot cancel $\omega_{q}^{\prime}$. Setting the left hand side of Eq.~(\ref{eq:dressed_mag_sense}) to zero gives transcendental equation. By taking the limit $\delta \gg \Omega_{\alpha}$, however, we can approximate the point where $\omega_{I,q}^{\prime}=0$. In this limit, a given dressing field strength will render the qubit subspace insensitive to magnetic fields when:
\begin{eqnarray}\label{eq:analytic_delta_clock}
    \delta = \sqrt{\frac{\omega^{\prime}_{z}(\Omega_{a}^2+\Omega_{b}^{2})}{4\omega_{q}^{\prime}}}.
\end{eqnarray}
We illustrate this technique in Fig.~\ref{fig:b_sense} for our example $^{137}\text{Ba}^{+}$ $S_{1/2}$ manifold, including the quadratic shift due to non-zero $B_{q}$. As shown, we can eliminate the the magnetic sensitivity of the $\{\ket{2,0},\ket{1,0}\}$ qubit subspace in a $2$~Gauss quantization field for multiple dressing field strengths: $10,~20,~30~\text{mG}$, giving AC shifts of approximately $\sim 300,~\sim 500,~\sim 800~\text{Hz}$, respectively. Here, we can see that the larger our dressing field amplitude, the smaller the $2^{\text{nd}}$-order magnetic sensitivity at the point when $\omega_{I,q}^{\prime}\simeq 0$\textemdash indicating stronger dressing fields offer more protection from noise. At the same time, larger dressing fields give larger shifts, increasing sensitivity to control noise. 

\subsection{Shelving}\label{sec:use_cases}

\begin{figure}[b]
\includegraphics[width=0.5\textwidth]{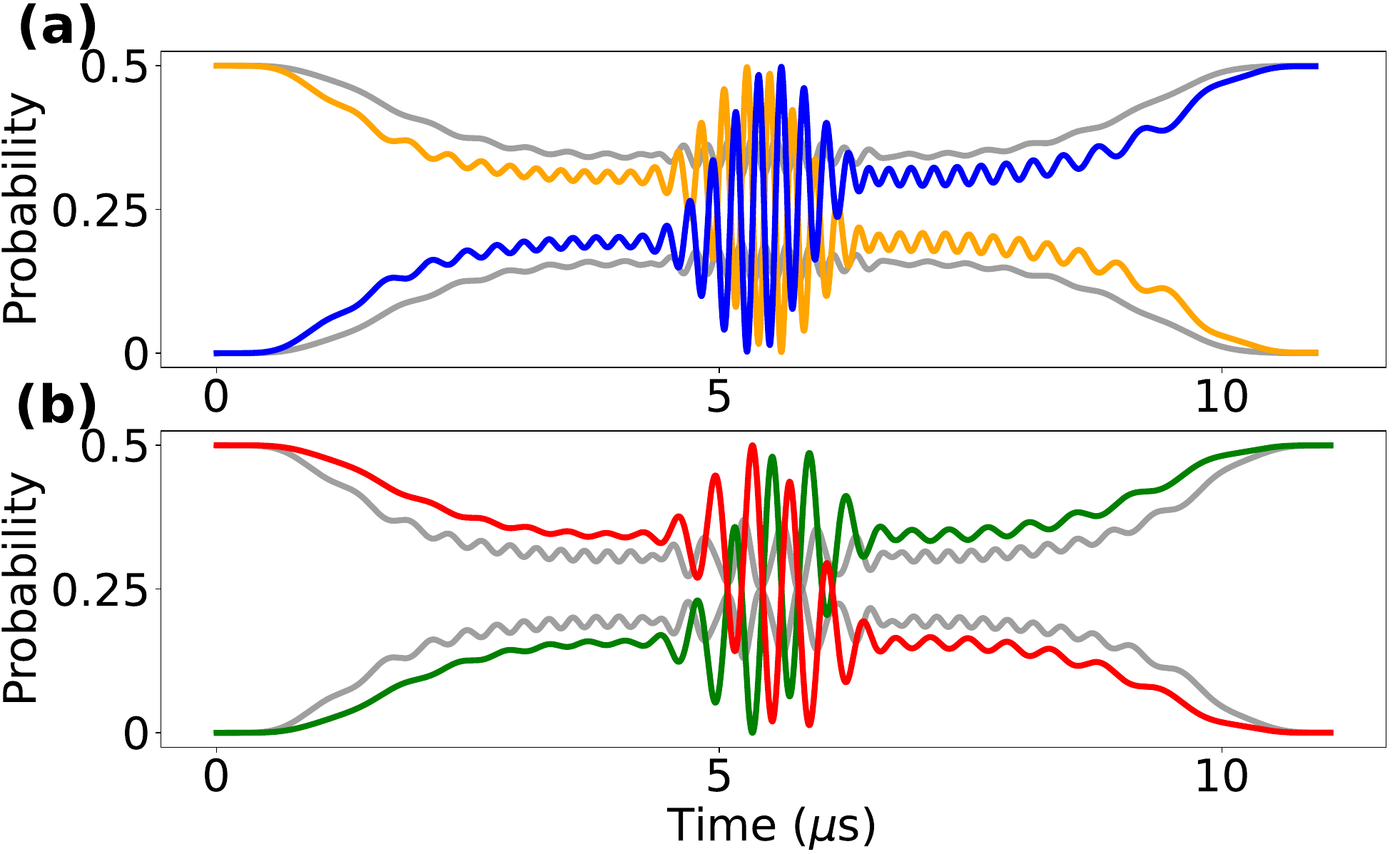}
\centering
\caption{Example of dressing field enabled shelving out of the $\{\ket{2,0},\ket{1,0}\}$ qubit manifold in the $S_{1/2}$ ground state of $^{137}\text{Ba}^{+}$. In both figures, we ramp on an $x$-polarized dressing field to $5~$Gauss over $4~\mu\text{s}$, with a detuning $\delta/2\pi \simeq 1~\text{MHz}$ from the $\ket{1,0}\leftrightarrow \ket{2,1}$ and $\ket{2,0}\leftrightarrow \ket{1,1}$ transitions. We then apply a second $\sim \text{MHz}$ frequency $z$-polarized field at the selected transition's dressed state splitting. (a) We set the frequency of the second field to $\Delta_{a}$, the dressed state splitting of the $\ket{1,0}\leftrightarrow\ket{2,1}$ transition, where the probability of $\ket{1,0}$ is in blue (bottom middle when $t\simeq 0$) and the probability of $\ket{2,1}$ is in orange (top middle when $t\simeq 0$). (b) We set the frequency of the second field to $\Delta_{b}$, the dressed state splitting of the $\ket{2,0}\leftrightarrow\ket{1,1}$ transition, where the probability of $\ket{2,0}$ is in red (top when $t\simeq 0$) and the probability of $\ket{1,1}$ is in green (bottom when $t\simeq 0$).}
\label{fig:shelve}
\end{figure}

Quadratic Zeeman shifts are small when operating at `low' quantization fields, leading to one of the regime's major drawbacks: that many transition pairs are nearly degenerate. This prevents frequency selection, which makes many potential shelving schemes unfeasible. Consider again the case of the $\{\ket{2,0},\ket{1,0}\}$ $S_{1/2}$ ground state qubit of $^{137}\text{Ba}^{+}$. It would be difficult to map this onto any other qubit pair because every $\Delta m_{F}=\pm 1$ transition is nearly-degenerate with another transition; the frequency of $\ket{2,0}\leftrightarrow \ket{1,1}$, for example, is degenerate with $\ket{1,0}\leftrightarrow \ket{2,1}$\textemdash up to the quadratic shift. It is possible, in principle, to select this transition with circularly polarized light, but it would be difficult to operate at high fidelity using lasers and circularly polarized near-field microwaves have yet to be demonstrated. In this section, we discuss a way around the issue that uses only linearly polarized microwaves. First, we ramp on the dressing field described by Eq.~(\ref{eq:h_mu}) to break the degeracy of the two transitions. Then, we tune the frequency of a second, $z$-polarized driving field to the splitting of one of the two dressed manifolds. By modulating the Zeeman shift, we add an on-resonant tone to the effective Hamiltonian of the system. After we adiabatically ramp off the dressing field, the populations of the targeted states will have exchanged (see Fig.~\ref{fig:shelve}).  \\

\noindent Consider a target Hamiltonian generated by a $z$-polarized magnetic field sinusoidally oscillating $\sim\text{MHz}$, which lets us ignore $\sim\text{GHz}$ transitions between the hyperfine manifolds. Written in the rotating frame with respect to the hyperfine $\omega_{0}$ and Zeeman $\omega_{z}$ splittings, this gives:
\begin{eqnarray}\label{eq:target_general}
    \hat{H}_{{t}} &= & \mu_{\text{B}}B_{{t}}(t)g_{J}\hat{J}_{z} \\    &=&\!\!\!\frac{\hbar}{2}\sum_{m_{F}^{\prime}}\Omega_{m_{F}^{\prime}}(t)\Big(\ket{F^{+},m_{F}^{\prime}}\!\bra{F^{+}\!\!,m_{F}^{\prime}}-\ket{F^{-}\!\!,m_{F}^{\prime}}\!\bra{F^{-}\!\!,m_{F}^{\prime}}\Big), \nonumber
\end{eqnarray}
where $\Omega_{m_{F}^{\prime}}(t)\rightarrow \Omega(t)m_{F}^{\prime}$, since we will, here, assume quadratic shifts are negligible. As discussed, we can diagonalize $\hat{H}_{D}$ to calculate the Rabi frequency and frequency splittings of the transitions we intend to drive. We can transform $\hat{H}_{{t}}$ into the dressed basis using Eq.~(\ref{eq:diag_u}). Written in the rotating frame with respect to the dressed state splitting, given by Eq.~(\ref{eq:dressed_sigmaz}), and making the rotating wave approximation with respect to the (off-resonant) diagonal terms gives:
\begin{eqnarray}
    \hat{H}_{I,\text{s}} \!\!\!&=&\!\!\! -\frac{\hbar\Omega(t)}{4}(2m_{F}\!\pm\! 1)\Big\{\!\!\sin(\theta_{a})\!\ket{F^{+}\!\!\!,m_{F}\!\pm\! 1}\!\!\bra{F^{-}\!\!,m_{F}}\!e^{i\Delta_{a}t} \nonumber \\
    && ~~~~~~~~~~~~~~~~~+\sin(\theta_{b})\!\ket{F^{+}\!\!\!,m_{F}}\!\!\bra{F^{-}\!\!,m_{F}\!\pm\! 1}\!e^{i\Delta_{b}t}\Big\} \nonumber \\
    &&~~~~~~~~~~~~~~~~~+ \text{c.c.} 
\end{eqnarray}
We can see here that the dressing field has broken the degeneracy, allowing us to now frequency select either transition by setting $\Omega(t)=\Omega_{0}\cos(\omega_{a}t)$ or $\Omega(t)=\Omega_{0}\cos(\omega_{b}t)$. \\

\noindent We demonstrate this in Fig.~\ref{fig:shelve}, using the technique to shelve out of our example $^{137}\text{Ba}^{+}$ $\{\ket{2,0},\ket{1,0}\}$ qubit manifold. To suppress off-resonant transitions, we ramp the dressing field on/off over a time $t_{D}$ according to a $\propto \sin^{2}(t\pi/t_{D})$ envelope. After/before, we apply a $\propto \sin^{2}(\pi t/t_{s})$ envelope to $\Omega(t)$ for the entire pulse duration $t_{s}$. In both figures, we apply the same dressing field. In Fig.~\ref{fig:shelve}a, we apply $\Delta_{a}$, which results in population transfer from $\ket{1,0}$ to $\ket{2,1}$, and no transfer from $\ket{2,0}$ to $\ket{1,1}$. In Fig.~\ref{fig:shelve}b, conversely, we apply $\Delta_{b}$, showing transfer from $\ket{2,0}$ to $\ket{1,1}$ and no transfer from $\ket{1,0}$ to $\ket{2,1}$. Since we have ignored the quadratic Zeeman shift, the transitions are exactly degenerate in the simulation, and it would not be possible to frequency select either without dressing. We can apply this general technique to any pair of (nearly) degenerate transitions in the ground state manifold\textemdash as long as $\Omega_{a}\neq \Omega_{b}$.

\subsection{Frequency separated single qubit gates}\label{sec:one_qubit}

\begin{figure}[b]
\includegraphics[width=0.5\textwidth]{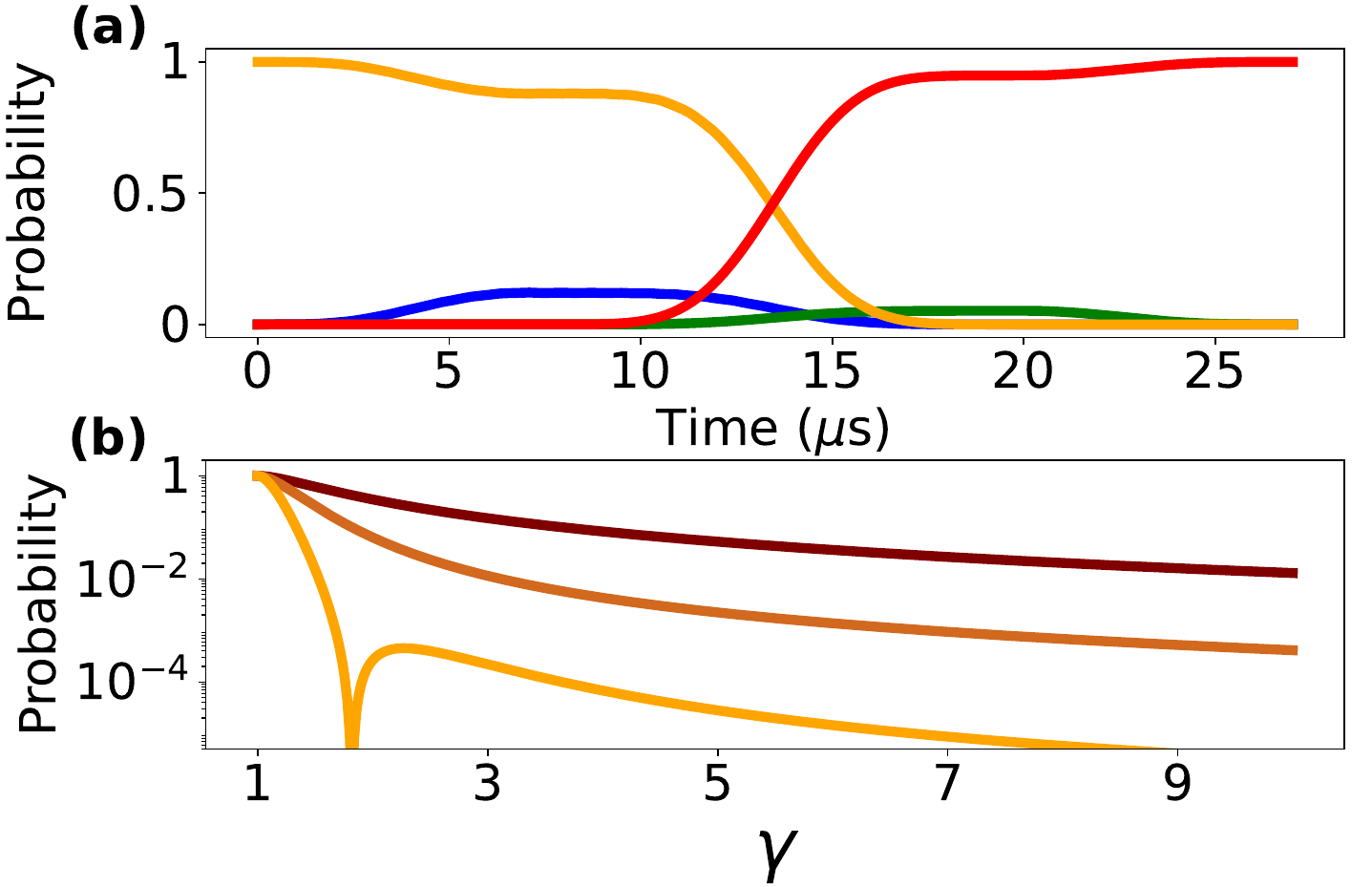}
\centering
\caption{Example of frequency separated single qubit gate for our example  $^{137}\text{Ba}^{+}$ $S_{1/2}$ ground state qubit manifold $\{\ket{2,0},\ket{1,0}\}$. We initialize the system to $\ket{\psi_{0}}=\ket{1,0}$. Probability of being in state $\ket{1,0}$ (orange, top near $t\simeq 0$), $\ket{2,0}$ (red, bottom near $t\simeq 0$), $\ket{2,1}$ (blue, reaches max near $t\sim 8~\mu\text{s}$), and $\ket{1,1}$ (green, reaches max near $t\sim 18~\mu\text{s}$). Here, we apply a $1~\text{G}$ dressing field, and a $30~\text{mG}$ gating field. (b) Demonstration of crosstalk suppression. We plot the probability of leakage out of the initial state $\ket{1,0}$ versus the `field ratio' $\gamma$, i.e. the ratio of the field magnitudes experienced a target qubit to the field magnitudes experienced by an idle qubit. When $\gamma = 1$, we apply a $30~\text{mG}$ driving field after/before a $1~\text{G}$ (orange bottom), $0.75~\text{G}$ (tan middle), and $0.5~\text{G}$ (brown top) dressing field, which we ramp on/off over $7~\mu\text{s}$.}
\label{fig:gate}
\end{figure}

Any single qubit gate scheme will need to operate with minimal crosstalk, preferably using a technique that does not complicate with system size. Frequency separation is a promising technique for this. In general, this works by implementing a gate scheme that requires only control frequencies that are far-detuned from any relevant transition in the system's `idle' qubits \cite{ospelkaus_2011,sutherland_2023,srinivas_2023}. Below, we discuss how to do this with dressing fields. \\

\noindent Keeping the pattern of the prior sections, we illustrate this using a qubit manifold $\{\ket{F^{+},m_{F}},\ket{F^{-},m_{F}}\}$, i.e. one that is directly addressable with a $z$-polarized magnetic field oscillating near $\omega_{0}$. This is given by:
\begin{eqnarray}
    \hat{H}_{t,g} \!&=& \!\!\mu_{\text{B}}B_{q}\cos\Big(\![\omega_{0}+\varepsilon]t \Big)\hat{J}_{z} \nonumber \\
    &\simeq & \!\!\!\!\frac{\hbar\Omega_{g}}{2}\Big(\!\ket{F^{+}\!,m_{F}}\!\bra{F^{-}\!,m_{F}}\!e^{-i\varepsilon t}\!\! +  \ket{F^{-}\!,m_{F}}\!\bra{F^{+}\!,m_{F}}e^{i\varepsilon t}\Big), \nonumber \\
\end{eqnarray}
where, we made the rotating wave approximation in the second line. We can rewrite this equation in a stationary frame by applying the transformation:
\begin{eqnarray}
    \hat{U}_{\varepsilon}\equiv \exp\Big(-\frac{i\varepsilon}{2}\hat{\sigma}_{z}\Big)
\end{eqnarray}
where $\hat{\sigma}_{\alpha}$ is a Pauli operator acting on the qubit manifold $\ket{F^{\pm},m_{F}}$. Projecting onto the qubit manifold gives:
\begin{eqnarray}
    \hat{H}_{t,g}^{\prime}&=& \frac{\hbar\varepsilon}{2}\hat{\sigma}_{z} +\frac{\hbar\Omega_{g}}{2}\hat{\sigma}_{x}.
\end{eqnarray}
To transform into the dressed basis, we can, again, use Eq.~(\ref{eq:diag_u}), which gives:
\begin{eqnarray}
    \hat{H}_{t,g}^{\prime} &=& \frac{\hbar}{2}\Big([\bar{\Delta}-\delta] -\varepsilon\Big)\hat{\sigma}_{z}+\frac{\hbar\Omega_{g}^{\prime}}{2}\hat{\sigma}_{x}, \nonumber \\
\end{eqnarray}
where $\Omega^{\prime}_{g}\equiv \cos\Big(\frac{\theta_{a}}{2}\Big)\cos\Big(\frac{\theta_{b}}{2}\Big)\Omega_{g}$ and $\bar{\Delta}\equiv (\Delta_{a}+\Delta_{b})/2$, indicating we have shifted the frequency resonant with the qubit transition by an amount:
\begin{eqnarray}
    \varepsilon &=& \bar{\Delta}-\delta.
\end{eqnarray}
Since the value of $\bar{\Delta}$ depends strongly on the ions' distance from the dressing wire, the value of $\varepsilon$ for idle qubits will be much different than for targeted qubits, giving frequency separation. \\

\noindent In Fig.~\ref{fig:gate}a, we illustrate a gate operation using the scheme for our example $S_{1/2}$ ground state manifold of $^{137}\text{Ba}^{+}$, again taking the qubit manifold to be $\{\ket{2,0},\ket{1,0}\}$. We initialize the system to $\ket{\psi(0)}=\ket{1,0}$, plotting the probability of each state versus time for a $\pi-$pulse when we apply a $1~\text{G}$ dressing field that we detune by $\delta/2\pi = 1~\text{MHz}$ and ramp on/off over $7~\mu\text{s}$, before/after we apply a $30~\text{mG}$ gating field. The value of $\varepsilon$ needed to drive the gate resonantly depends on the $\textit{size}$ of the dressing field. This means that an `idle' ion, assumed to be much further away from the gating wire than the target ion, will experience only off-resonant magnetic fields, suppressing crosstalk. We can model this (roughly) by uniformly dividing the magnitude of both the dressing and gating fields in our  simulations by a constant value $\gamma$ (i.e. the dressing and driving magnetic fields are made a factor of $\gamma$ smaller than their ideal values) and analyze the probability of the gating fields changing the state of the ion. There will of course be an AC shift associated with the idle qubits, but this can be accounted for \cite{pino_2021}. In Fig.~\ref{fig:gate}b, we plot the probability of state leakage out of an initial state $\ket{1,0}$, i.e. $1-|\braket{\psi(t)|1,0}|^{2}$, versus the value of $\gamma$. Note, we do not change the frequencies of any fields. For each plot, we use a $30~\text{mG}$ driving field, showing the results for $0.5, 0.75$, and $1~\text{Gauss}$ dressing fields. For the latter field size, can see the probability of state leakage is suppressed well-below $10^{-5}$ when $\gamma \gtrsim 10$. Assuming an ion distance of $\sim 50~\mu\text{m}$, this means that once an idle qubit is $\sim 500~\mu\text{m}$ away from the gating zone, the main effect our gating fields have is a traceable AC shift, decreasing quadratically in field size. \\

\subsection{Conclusion}\label{sec:conclusion}

In this work, we discuss how a dressing field delivered via the near field of an integrated wire can manipulate the hyperfine structure of any atom near the wire and propose three use cases. The first is to eliminate small linear magnetic sensitivities in a qubit, potentially increasing memory times. The the second is to break the degeneracies that often make shelving difficult when using small quantization fields. This could allow experimentalists to map between arbitrary pairs of qubit sublevels, enabling many experiments that otherwise would not be possible at $\lesssim 10~\text{Gauss}$ quantization fields. Last, we discussed a method to frequency separate single qubit gates implemented on `target' qubits from `idle' ions, not in a specified location.

\section*{Acknowledgements}

We would like to thank M. Foss-Feig, J. Gaebler, B. J. Bjork, C. Langer, M. Schecter, J. Bartolotta, and P. J. Lee for helpful discussions, as well as D. T. C. Allcock and R. Matt for reviewing the manuscript.

\bibliography{biblio}

\begin{thebibliography}{28}%
\makeatletter
\providecommand \@ifxundefined [1]{%
 \@ifx{#1\undefined}
}%
\providecommand \@ifnum [1]{%
 \ifnum #1\expandafter \@firstoftwo
 \else \expandafter \@secondoftwo
 \fi
}%
\providecommand \@ifx [1]{%
 \ifx #1\expandafter \@firstoftwo
 \else \expandafter \@secondoftwo
 \fi
}%
\providecommand \natexlab [1]{#1}%
\providecommand \enquote  [1]{``#1''}%
\providecommand \bibnamefont  [1]{#1}%
\providecommand \bibfnamefont [1]{#1}%
\providecommand \citenamefont [1]{#1}%
\providecommand \href@noop [0]{\@secondoftwo}%
\providecommand \href [0]{\begingroup \@sanitize@url \@href}%
\providecommand \@href[1]{\@@startlink{#1}\@@href}%
\providecommand \@@href[1]{\endgroup#1\@@endlink}%
\providecommand \@sanitize@url [0]{\catcode `\\12\catcode `\$12\catcode
  `\&12\catcode `\#12\catcode `\^12\catcode `\_12\catcode `\%12\relax}%
\providecommand \@@startlink[1]{}%
\providecommand \@@endlink[0]{}%
\providecommand \url  [0]{\begingroup\@sanitize@url \@url }%
\providecommand \@url [1]{\endgroup\@href {#1}{\urlprefix }}%
\providecommand \urlprefix  [0]{URL }%
\providecommand \Eprint [0]{\href }%
\providecommand \doibase [0]{http://dx.doi.org/}%
\providecommand \selectlanguage [0]{\@gobble}%
\providecommand \bibinfo  [0]{\@secondoftwo}%
\providecommand \bibfield  [0]{\@secondoftwo}%
\providecommand \translation [1]{[#1]}%
\providecommand \BibitemOpen [0]{}%
\providecommand \bibitemStop [0]{}%
\providecommand \bibitemNoStop [0]{.\EOS\space}%
\providecommand \EOS [0]{\spacefactor3000\relax}%
\providecommand \BibitemShut  [1]{\csname bibitem#1\endcsname}%
\let\auto@bib@innerbib\@empty
%</preamble>
\bibitem [{\citenamefont {Wineland}\ \emph {et~al.}(1998)\citenamefont
  {Wineland}, \citenamefont {Monroe}, \citenamefont {Itano}, \citenamefont
  {Leibfried}, \citenamefont {King},\ and\ \citenamefont
  {Meekhof}}]{wineland_1998}%
  \BibitemOpen
  \bibfield  {author} {\bibinfo {author} {\bibfnamefont {D.~J.}\ \bibnamefont
  {Wineland}}, \bibinfo {author} {\bibfnamefont {C.}~\bibnamefont {Monroe}},
  \bibinfo {author} {\bibfnamefont {W.~M.}\ \bibnamefont {Itano}}, \bibinfo
  {author} {\bibfnamefont {D.}~\bibnamefont {Leibfried}}, \bibinfo {author}
  {\bibfnamefont {B.~E.}\ \bibnamefont {King}}, \ and\ \bibinfo {author}
  {\bibfnamefont {D.~M.}\ \bibnamefont {Meekhof}},\ }\href@noop {} {\bibfield
  {journal} {\bibinfo  {journal} {J. Res. Natl. Inst. Stand. and Technol.}\
  }\textbf {\bibinfo {volume} {103}},\ \bibinfo {pages} {259} (\bibinfo {year}
  {1998})}\BibitemShut {NoStop}%
\bibitem [{\citenamefont {Ladd}\ \emph {et~al.}(2010)\citenamefont {Ladd},
  \citenamefont {Jelezko}, \citenamefont {Laflamme}, \citenamefont {Nakamura},
  \citenamefont {Monroe},\ and\ \citenamefont {O'Brien}}]{ladd_2010}%
  \BibitemOpen
  \bibfield  {author} {\bibinfo {author} {\bibfnamefont {T.~D.}\ \bibnamefont
  {Ladd}}, \bibinfo {author} {\bibfnamefont {F.}~\bibnamefont {Jelezko}},
  \bibinfo {author} {\bibfnamefont {R.}~\bibnamefont {Laflamme}}, \bibinfo
  {author} {\bibfnamefont {Y.}~\bibnamefont {Nakamura}}, \bibinfo {author}
  {\bibfnamefont {C.}~\bibnamefont {Monroe}}, \ and\ \bibinfo {author}
  {\bibfnamefont {J.~L.}\ \bibnamefont {O'Brien}},\ }\href@noop {} {\bibfield
  {journal} {\bibinfo  {journal} {Nature}\ }\textbf {\bibinfo {volume} {464}},\
  \bibinfo {pages} {45} (\bibinfo {year} {2010})}\BibitemShut {NoStop}%
\bibitem [{\citenamefont {Harty}\ \emph {et~al.}(2014)\citenamefont {Harty},
  \citenamefont {Allcock}, \citenamefont {Ballance}, \citenamefont {Guidoni},
  \citenamefont {Janacek}, \citenamefont {Linke}, \citenamefont {Stacey},\ and\
  \citenamefont {Lucas}}]{harty_2014}%
  \BibitemOpen
  \bibfield  {author} {\bibinfo {author} {\bibfnamefont {T.~P.}\ \bibnamefont
  {Harty}}, \bibinfo {author} {\bibfnamefont {D.~T.~C.}\ \bibnamefont
  {Allcock}}, \bibinfo {author} {\bibfnamefont {C.~J.}\ \bibnamefont
  {Ballance}}, \bibinfo {author} {\bibfnamefont {L.}~\bibnamefont {Guidoni}},
  \bibinfo {author} {\bibfnamefont {H.~A.}\ \bibnamefont {Janacek}}, \bibinfo
  {author} {\bibfnamefont {N.~M.}\ \bibnamefont {Linke}}, \bibinfo {author}
  {\bibfnamefont {D.~N.}\ \bibnamefont {Stacey}}, \ and\ \bibinfo {author}
  {\bibfnamefont {D.~M.}\ \bibnamefont {Lucas}},\ }\href {\doibase
  10.1103/PhysRevLett.113.220501} {\bibfield  {journal} {\bibinfo  {journal}
  {Phys. Rev. Lett.}\ }\textbf {\bibinfo {volume} {113}},\ \bibinfo {pages}
  {220501} (\bibinfo {year} {2014})}\BibitemShut {NoStop}%
\bibitem [{\citenamefont {Ballance}\ \emph {et~al.}(2016)\citenamefont
  {Ballance}, \citenamefont {Harty}, \citenamefont {Linke}, \citenamefont
  {Sepiol},\ and\ \citenamefont {Lucas}}]{ballance_2016}%
  \BibitemOpen
  \bibfield  {author} {\bibinfo {author} {\bibfnamefont {C.~J.}\ \bibnamefont
  {Ballance}}, \bibinfo {author} {\bibfnamefont {T.~P.}\ \bibnamefont {Harty}},
  \bibinfo {author} {\bibfnamefont {N.~M.}\ \bibnamefont {Linke}}, \bibinfo
  {author} {\bibfnamefont {M.~A.}\ \bibnamefont {Sepiol}}, \ and\ \bibinfo
  {author} {\bibfnamefont {D.~M.}\ \bibnamefont {Lucas}},\ }\href {\doibase
  10.1103/PhysRevLett.117.060504} {\bibfield  {journal} {\bibinfo  {journal}
  {Phys. Rev. Lett.}\ }\textbf {\bibinfo {volume} {117}},\ \bibinfo {pages}
  {060504} (\bibinfo {year} {2016})}\BibitemShut {NoStop}%
\bibitem [{\citenamefont {Gaebler}\ \emph {et~al.}(2016)\citenamefont
  {Gaebler}, \citenamefont {Tan}, \citenamefont {Lin}, \citenamefont {Wan},
  \citenamefont {Bowler}, \citenamefont {Keith}, \citenamefont {Glancy},
  \citenamefont {Coakley}, \citenamefont {Knill}, \citenamefont {Leibfried},\
  and\ \citenamefont {Wineland}}]{gaebler_2016}%
  \BibitemOpen
  \bibfield  {author} {\bibinfo {author} {\bibfnamefont {J.~P.}\ \bibnamefont
  {Gaebler}}, \bibinfo {author} {\bibfnamefont {T.~R.}\ \bibnamefont {Tan}},
  \bibinfo {author} {\bibfnamefont {Y.}~\bibnamefont {Lin}}, \bibinfo {author}
  {\bibfnamefont {Y.}~\bibnamefont {Wan}}, \bibinfo {author} {\bibfnamefont
  {R.}~\bibnamefont {Bowler}}, \bibinfo {author} {\bibfnamefont {A.~C.}\
  \bibnamefont {Keith}}, \bibinfo {author} {\bibfnamefont {S.}~\bibnamefont
  {Glancy}}, \bibinfo {author} {\bibfnamefont {K.}~\bibnamefont {Coakley}},
  \bibinfo {author} {\bibfnamefont {E.}~\bibnamefont {Knill}}, \bibinfo
  {author} {\bibfnamefont {D.}~\bibnamefont {Leibfried}}, \ and\ \bibinfo
  {author} {\bibfnamefont {D.~J.}\ \bibnamefont {Wineland}},\ }\href {\doibase
  10.1103/PhysRevLett.117.060505} {\bibfield  {journal} {\bibinfo  {journal}
  {Phys. Rev. Lett.}\ }\textbf {\bibinfo {volume} {117}},\ \bibinfo {pages}
  {060505} (\bibinfo {year} {2016})}\BibitemShut {NoStop}%
\bibitem [{\citenamefont {Srinivas}\ \emph {et~al.}(2021)\citenamefont
  {Srinivas}, \citenamefont {Burd}, \citenamefont {Knaack}, \citenamefont
  {Sutherland}, \citenamefont {Kwiatkowski}, \citenamefont {Glancy},
  \citenamefont {Knill}, \citenamefont {Wineland}, \citenamefont {Leibfried},
  \citenamefont {Wilson}, \citenamefont {C.},\ and\ \citenamefont
  {H.}}]{srinivas_2021}%
  \BibitemOpen
  \bibfield  {author} {\bibinfo {author} {\bibfnamefont {R.}~\bibnamefont
  {Srinivas}}, \bibinfo {author} {\bibfnamefont {S.~C.}\ \bibnamefont {Burd}},
  \bibinfo {author} {\bibfnamefont {H.~M.}\ \bibnamefont {Knaack}}, \bibinfo
  {author} {\bibfnamefont {R.~T.}\ \bibnamefont {Sutherland}}, \bibinfo
  {author} {\bibfnamefont {A.}~\bibnamefont {Kwiatkowski}}, \bibinfo {author}
  {\bibfnamefont {S.}~\bibnamefont {Glancy}}, \bibinfo {author} {\bibfnamefont
  {E.}~\bibnamefont {Knill}}, \bibinfo {author} {\bibfnamefont {D.~J.}\
  \bibnamefont {Wineland}}, \bibinfo {author} {\bibfnamefont {D.}~\bibnamefont
  {Leibfried}}, \bibinfo {author} {\bibfnamefont {A.~C.}\ \bibnamefont
  {Wilson}}, \bibinfo {author} {\bibfnamefont {A.~D.~T.}\ \bibnamefont {C.}}, \
  and\ \bibinfo {author} {\bibfnamefont {S.~D.}\ \bibnamefont {H.}},\
  }\href@noop {} {\bibfield  {journal} {\bibinfo  {journal} {Nature}\ }\textbf
  {\bibinfo {volume} {597}},\ \bibinfo {pages} {209} (\bibinfo {year}
  {2021})}\BibitemShut {NoStop}%
\bibitem [{\citenamefont {Pino}\ \emph {et~al.}(2021)\citenamefont {Pino},
  \citenamefont {Dreiling}, \citenamefont {Figgatt}, \citenamefont {Gaebler},
  \citenamefont {Moses}, \citenamefont {Allman}, \citenamefont {Baldwin},
  \citenamefont {Foss-Feig}, \citenamefont {Hayes}, \citenamefont {Mayer} \emph
  {et~al.}}]{pino_2021}%
  \BibitemOpen
  \bibfield  {author} {\bibinfo {author} {\bibfnamefont {J.~M.}\ \bibnamefont
  {Pino}}, \bibinfo {author} {\bibfnamefont {J.~M.}\ \bibnamefont {Dreiling}},
  \bibinfo {author} {\bibfnamefont {C.}~\bibnamefont {Figgatt}}, \bibinfo
  {author} {\bibfnamefont {J.~P.}\ \bibnamefont {Gaebler}}, \bibinfo {author}
  {\bibfnamefont {S.~A.}\ \bibnamefont {Moses}}, \bibinfo {author}
  {\bibfnamefont {M.}~\bibnamefont {Allman}}, \bibinfo {author} {\bibfnamefont
  {C.}~\bibnamefont {Baldwin}}, \bibinfo {author} {\bibfnamefont
  {M.}~\bibnamefont {Foss-Feig}}, \bibinfo {author} {\bibfnamefont
  {D.}~\bibnamefont {Hayes}}, \bibinfo {author} {\bibfnamefont
  {K.}~\bibnamefont {Mayer}},  \emph {et~al.},\ }\href@noop {} {\bibfield
  {journal} {\bibinfo  {journal} {Nature}\ }\textbf {\bibinfo {volume} {592}},\
  \bibinfo {pages} {209} (\bibinfo {year} {2021})}\BibitemShut {NoStop}%
\bibitem [{\citenamefont {Malinowski}\ \emph {et~al.}(2023)\citenamefont
  {Malinowski}, \citenamefont {Allcock},\ and\ \citenamefont
  {Ballance}}]{malinowski_2023}%
  \BibitemOpen
  \bibfield  {author} {\bibinfo {author} {\bibfnamefont {M.}~\bibnamefont
  {Malinowski}}, \bibinfo {author} {\bibfnamefont {D.}~\bibnamefont {Allcock}},
  \ and\ \bibinfo {author} {\bibfnamefont {C.}~\bibnamefont {Ballance}},\
  }\href@noop {} {\bibfield  {journal} {\bibinfo  {journal} {PRX Quantum}\
  }\textbf {\bibinfo {volume} {4}},\ \bibinfo {pages} {040313} (\bibinfo {year}
  {2023})}\BibitemShut {NoStop}%
\bibitem [{\citenamefont {Moses}\ \emph {et~al.}(2023)\citenamefont {Moses},
  \citenamefont {Baldwin}, \citenamefont {Allman}, \citenamefont {Ancona},
  \citenamefont {Ascarrunz}, \citenamefont {Barnes}, \citenamefont
  {Bartolotta}, \citenamefont {Bjork}, \citenamefont {Blanchard}, \citenamefont
  {Bohn} \emph {et~al.}}]{moses_2023}%
  \BibitemOpen
  \bibfield  {author} {\bibinfo {author} {\bibfnamefont {S.~A.}\ \bibnamefont
  {Moses}}, \bibinfo {author} {\bibfnamefont {C.~H.}\ \bibnamefont {Baldwin}},
  \bibinfo {author} {\bibfnamefont {M.~S.}\ \bibnamefont {Allman}}, \bibinfo
  {author} {\bibfnamefont {R.}~\bibnamefont {Ancona}}, \bibinfo {author}
  {\bibfnamefont {L.}~\bibnamefont {Ascarrunz}}, \bibinfo {author}
  {\bibfnamefont {C.}~\bibnamefont {Barnes}}, \bibinfo {author} {\bibfnamefont
  {J.}~\bibnamefont {Bartolotta}}, \bibinfo {author} {\bibfnamefont
  {B.}~\bibnamefont {Bjork}}, \bibinfo {author} {\bibfnamefont
  {P.}~\bibnamefont {Blanchard}}, \bibinfo {author} {\bibfnamefont
  {M.}~\bibnamefont {Bohn}},  \emph {et~al.},\ }\href@noop {} {\bibfield
  {journal} {\bibinfo  {journal} {Physical Review X}\ }\textbf {\bibinfo
  {volume} {13}},\ \bibinfo {pages} {041052} (\bibinfo {year}
  {2023})}\BibitemShut {NoStop}%
\bibitem [{\citenamefont {Harty}\ \emph {et~al.}(2016)\citenamefont {Harty},
  \citenamefont {Sepiol}, \citenamefont {Allcock}, \citenamefont {Ballance},
  \citenamefont {Tarlton},\ and\ \citenamefont {Lucas}}]{harty_2016}%
  \BibitemOpen
  \bibfield  {author} {\bibinfo {author} {\bibfnamefont {T.~P.}\ \bibnamefont
  {Harty}}, \bibinfo {author} {\bibfnamefont {M.~A.}\ \bibnamefont {Sepiol}},
  \bibinfo {author} {\bibfnamefont {D.~T.~C.}\ \bibnamefont {Allcock}},
  \bibinfo {author} {\bibfnamefont {C.~J.}\ \bibnamefont {Ballance}}, \bibinfo
  {author} {\bibfnamefont {J.~E.}\ \bibnamefont {Tarlton}}, \ and\ \bibinfo
  {author} {\bibfnamefont {D.~M.}\ \bibnamefont {Lucas}},\ }\href@noop {}
  {\bibfield  {journal} {\bibinfo  {journal} {Phys. Rev. Lett.}\ }\textbf
  {\bibinfo {volume} {117}},\ \bibinfo {pages} {140501} (\bibinfo {year}
  {2016})}\BibitemShut {NoStop}%
\bibitem [{\citenamefont {Aude~Craik}\ \emph {et~al.}(2014)\citenamefont
  {Aude~Craik}, \citenamefont {Linke}, \citenamefont {Harty}, \citenamefont
  {Ballance}, \citenamefont {Lucas}, \citenamefont {Steane},\ and\
  \citenamefont {Allcock}}]{aude_2014}%
  \BibitemOpen
  \bibfield  {author} {\bibinfo {author} {\bibfnamefont {D.}~\bibnamefont
  {Aude~Craik}}, \bibinfo {author} {\bibfnamefont {N.}~\bibnamefont {Linke}},
  \bibinfo {author} {\bibfnamefont {T.}~\bibnamefont {Harty}}, \bibinfo
  {author} {\bibfnamefont {C.}~\bibnamefont {Ballance}}, \bibinfo {author}
  {\bibfnamefont {D.}~\bibnamefont {Lucas}}, \bibinfo {author} {\bibfnamefont
  {A.}~\bibnamefont {Steane}}, \ and\ \bibinfo {author} {\bibfnamefont
  {D.}~\bibnamefont {Allcock}},\ }\href@noop {} {\bibfield  {journal} {\bibinfo
   {journal} {Applied Physics B}\ }\textbf {\bibinfo {volume} {114}},\ \bibinfo
  {pages} {3} (\bibinfo {year} {2014})}\BibitemShut {NoStop}%
\bibitem [{\citenamefont {Aude~Craik}\ \emph {et~al.}(2017)\citenamefont
  {Aude~Craik}, \citenamefont {Linke}, \citenamefont {Sepiol}, \citenamefont
  {Harty}, \citenamefont {Goodwin}, \citenamefont {Ballance}, \citenamefont
  {Stacey}, \citenamefont {Steane}, \citenamefont {Lucas},\ and\ \citenamefont
  {Allcock}}]{aude_2017}%
  \BibitemOpen
  \bibfield  {author} {\bibinfo {author} {\bibfnamefont {D.~P.~L.}\
  \bibnamefont {Aude~Craik}}, \bibinfo {author} {\bibfnamefont {N.~M.}\
  \bibnamefont {Linke}}, \bibinfo {author} {\bibfnamefont {M.~A.}\ \bibnamefont
  {Sepiol}}, \bibinfo {author} {\bibfnamefont {T.~P.}\ \bibnamefont {Harty}},
  \bibinfo {author} {\bibfnamefont {J.~F.}\ \bibnamefont {Goodwin}}, \bibinfo
  {author} {\bibfnamefont {C.~J.}\ \bibnamefont {Ballance}}, \bibinfo {author}
  {\bibfnamefont {D.~N.}\ \bibnamefont {Stacey}}, \bibinfo {author}
  {\bibfnamefont {A.~M.}\ \bibnamefont {Steane}}, \bibinfo {author}
  {\bibfnamefont {D.~M.}\ \bibnamefont {Lucas}}, \ and\ \bibinfo {author}
  {\bibfnamefont {D.~T.~C.}\ \bibnamefont {Allcock}},\ }\href {\doibase
  10.1103/PhysRevA.95.022337} {\bibfield  {journal} {\bibinfo  {journal} {Phys.
  Rev. A}\ }\textbf {\bibinfo {volume} {95}},\ \bibinfo {pages} {022337}
  (\bibinfo {year} {2017})}\BibitemShut {NoStop}%
\bibitem [{\citenamefont {Leu}\ \emph {et~al.}(2023)\citenamefont {Leu},
  \citenamefont {Gely}, \citenamefont {Weber}, \citenamefont {Smith},
  \citenamefont {Nadlinger},\ and\ \citenamefont {Lucas}}]{leu_2023}%
  \BibitemOpen
  \bibfield  {author} {\bibinfo {author} {\bibfnamefont {A.}~\bibnamefont
  {Leu}}, \bibinfo {author} {\bibfnamefont {M.}~\bibnamefont {Gely}}, \bibinfo
  {author} {\bibfnamefont {M.}~\bibnamefont {Weber}}, \bibinfo {author}
  {\bibfnamefont {M.}~\bibnamefont {Smith}}, \bibinfo {author} {\bibfnamefont
  {D.}~\bibnamefont {Nadlinger}}, \ and\ \bibinfo {author} {\bibfnamefont
  {D.}~\bibnamefont {Lucas}},\ }\href@noop {} {\bibfield  {journal} {\bibinfo
  {journal} {Physical Review Letters}\ }\textbf {\bibinfo {volume} {131}},\
  \bibinfo {pages} {120601} (\bibinfo {year} {2023})}\BibitemShut {NoStop}%
\bibitem [{\citenamefont {Wang}\ \emph {et~al.}(2009)\citenamefont {Wang},
  \citenamefont {Labaziewicz}, \citenamefont {Ge}, \citenamefont {Shewmon},\
  and\ \citenamefont {Chuang}}]{wang_2009}%
  \BibitemOpen
  \bibfield  {author} {\bibinfo {author} {\bibfnamefont {S.~X.}\ \bibnamefont
  {Wang}}, \bibinfo {author} {\bibfnamefont {J.}~\bibnamefont {Labaziewicz}},
  \bibinfo {author} {\bibfnamefont {Y.}~\bibnamefont {Ge}}, \bibinfo {author}
  {\bibfnamefont {R.}~\bibnamefont {Shewmon}}, \ and\ \bibinfo {author}
  {\bibfnamefont {I.~L.}\ \bibnamefont {Chuang}},\ }\href@noop {} {\bibfield
  {journal} {\bibinfo  {journal} {Applied Physics Letters}\ }\textbf {\bibinfo
  {volume} {94}},\ \bibinfo {pages} {094103} (\bibinfo {year}
  {2009})}\BibitemShut {NoStop}%
\bibitem [{\citenamefont {Johanning}\ \emph {et~al.}(2009)\citenamefont
  {Johanning}, \citenamefont {Braun}, \citenamefont {Timoney}, \citenamefont
  {Elman}, \citenamefont {Neuhauser},\ and\ \citenamefont
  {Wunderlich}}]{johanning_2009}%
  \BibitemOpen
  \bibfield  {author} {\bibinfo {author} {\bibfnamefont {M.}~\bibnamefont
  {Johanning}}, \bibinfo {author} {\bibfnamefont {A.}~\bibnamefont {Braun}},
  \bibinfo {author} {\bibfnamefont {N.}~\bibnamefont {Timoney}}, \bibinfo
  {author} {\bibfnamefont {V.}~\bibnamefont {Elman}}, \bibinfo {author}
  {\bibfnamefont {W.}~\bibnamefont {Neuhauser}}, \ and\ \bibinfo {author}
  {\bibfnamefont {C.}~\bibnamefont {Wunderlich}},\ }\href {\doibase
  10.1103/PhysRevLett.102.073004} {\bibfield  {journal} {\bibinfo  {journal}
  {Phys. Rev. Lett.}\ }\textbf {\bibinfo {volume} {102}},\ \bibinfo {pages}
  {073004} (\bibinfo {year} {2009})}\BibitemShut {NoStop}%
\bibitem [{\citenamefont {Warring}\ \emph {et~al.}(2013)\citenamefont
  {Warring}, \citenamefont {Ospelkaus}, \citenamefont {Colombe}, \citenamefont
  {J\"ordens}, \citenamefont {Leibfried},\ and\ \citenamefont
  {Wineland}}]{warring_2013_prl}%
  \BibitemOpen
  \bibfield  {author} {\bibinfo {author} {\bibfnamefont {U.}~\bibnamefont
  {Warring}}, \bibinfo {author} {\bibfnamefont {C.}~\bibnamefont {Ospelkaus}},
  \bibinfo {author} {\bibfnamefont {Y.}~\bibnamefont {Colombe}}, \bibinfo
  {author} {\bibfnamefont {R.}~\bibnamefont {J\"ordens}}, \bibinfo {author}
  {\bibfnamefont {D.}~\bibnamefont {Leibfried}}, \ and\ \bibinfo {author}
  {\bibfnamefont {D.~J.}\ \bibnamefont {Wineland}},\ }\href {\doibase
  10.1103/PhysRevLett.110.173002} {\bibfield  {journal} {\bibinfo  {journal}
  {Phys. Rev. Lett.}\ }\textbf {\bibinfo {volume} {110}},\ \bibinfo {pages}
  {173002} (\bibinfo {year} {2013})}\BibitemShut {NoStop}%
\bibitem [{\citenamefont {Weidt}\ \emph {et~al.}(2016)\citenamefont {Weidt},
  \citenamefont {Randall}, \citenamefont {Webster}, \citenamefont {Lake},
  \citenamefont {Webb}, \citenamefont {Cohen}, \citenamefont {Navickas},
  \citenamefont {Lekitsch}, \citenamefont {Retzker},\ and\ \citenamefont
  {Hensinger}}]{weidt_2016}%
  \BibitemOpen
  \bibfield  {author} {\bibinfo {author} {\bibfnamefont {S.}~\bibnamefont
  {Weidt}}, \bibinfo {author} {\bibfnamefont {J.}~\bibnamefont {Randall}},
  \bibinfo {author} {\bibfnamefont {S.~C.}\ \bibnamefont {Webster}}, \bibinfo
  {author} {\bibfnamefont {K.}~\bibnamefont {Lake}}, \bibinfo {author}
  {\bibfnamefont {A.~E.}\ \bibnamefont {Webb}}, \bibinfo {author}
  {\bibfnamefont {I.}~\bibnamefont {Cohen}}, \bibinfo {author} {\bibfnamefont
  {T.}~\bibnamefont {Navickas}}, \bibinfo {author} {\bibfnamefont
  {B.}~\bibnamefont {Lekitsch}}, \bibinfo {author} {\bibfnamefont
  {A.}~\bibnamefont {Retzker}}, \ and\ \bibinfo {author} {\bibfnamefont
  {W.~K.}\ \bibnamefont {Hensinger}},\ }\href {\doibase
  10.1103/PhysRevLett.117.220501} {\bibfield  {journal} {\bibinfo  {journal}
  {Phys. Rev. Lett.}\ }\textbf {\bibinfo {volume} {117}},\ \bibinfo {pages}
  {220501} (\bibinfo {year} {2016})}\BibitemShut {NoStop}%
\bibitem [{\citenamefont {Sutherland}\ \emph {et~al.}(2023)\citenamefont
  {Sutherland}, \citenamefont {Srinivas},\ and\ \citenamefont
  {Allcock}}]{sutherland_2023}%
  \BibitemOpen
  \bibfield  {author} {\bibinfo {author} {\bibfnamefont {R.}~\bibnamefont
  {Sutherland}}, \bibinfo {author} {\bibfnamefont {R.}~\bibnamefont
  {Srinivas}}, \ and\ \bibinfo {author} {\bibfnamefont {D.}~\bibnamefont
  {Allcock}},\ }\href@noop {} {\bibfield  {journal} {\bibinfo  {journal}
  {Physical Review A}\ }\textbf {\bibinfo {volume} {107}},\ \bibinfo {pages}
  {032604} (\bibinfo {year} {2023})}\BibitemShut {NoStop}%
\bibitem [{\citenamefont {Ospelkaus}\ \emph {et~al.}(2008)\citenamefont
  {Ospelkaus}, \citenamefont {Langer}, \citenamefont {Amini}, \citenamefont
  {Brown}, \citenamefont {Leibfried},\ and\ \citenamefont
  {Wineland}}]{ospelkaus_2008}%
  \BibitemOpen
  \bibfield  {author} {\bibinfo {author} {\bibfnamefont {C.}~\bibnamefont
  {Ospelkaus}}, \bibinfo {author} {\bibfnamefont {C.~E.}\ \bibnamefont
  {Langer}}, \bibinfo {author} {\bibfnamefont {J.~M.}\ \bibnamefont {Amini}},
  \bibinfo {author} {\bibfnamefont {K.~R.}\ \bibnamefont {Brown}}, \bibinfo
  {author} {\bibfnamefont {D.}~\bibnamefont {Leibfried}}, \ and\ \bibinfo
  {author} {\bibfnamefont {D.~J.}\ \bibnamefont {Wineland}},\ }\href@noop {}
  {\bibfield  {journal} {\bibinfo  {journal} {Phys. Rev. Lett.}\ }\textbf
  {\bibinfo {volume} {101}},\ \bibinfo {pages} {090502} (\bibinfo {year}
  {2008})}\BibitemShut {NoStop}%
\bibitem [{\citenamefont {Ospelkaus}\ \emph {et~al.}(2011)\citenamefont
  {Ospelkaus}, \citenamefont {Warring}, \citenamefont {Colombe}, \citenamefont
  {Brown}, \citenamefont {Amini}, \citenamefont {Leibfried},\ and\
  \citenamefont {Wineland}}]{ospelkaus_2011}%
  \BibitemOpen
  \bibfield  {author} {\bibinfo {author} {\bibfnamefont {C.}~\bibnamefont
  {Ospelkaus}}, \bibinfo {author} {\bibfnamefont {U.}~\bibnamefont {Warring}},
  \bibinfo {author} {\bibfnamefont {Y.}~\bibnamefont {Colombe}}, \bibinfo
  {author} {\bibfnamefont {K.~R.}\ \bibnamefont {Brown}}, \bibinfo {author}
  {\bibfnamefont {J.~M.}\ \bibnamefont {Amini}}, \bibinfo {author}
  {\bibfnamefont {D.}~\bibnamefont {Leibfried}}, \ and\ \bibinfo {author}
  {\bibfnamefont {D.~J.}\ \bibnamefont {Wineland}},\ }\href@noop {} {\bibfield
  {journal} {\bibinfo  {journal} {Nature}\ }\textbf {\bibinfo {volume} {476}},\
  \bibinfo {pages} {181} (\bibinfo {year} {2011})}\BibitemShut {NoStop}%
\bibitem [{\citenamefont {Srinivas}\ \emph {et~al.}(2019)\citenamefont
  {Srinivas}, \citenamefont {Burd}, \citenamefont {Sutherland}, \citenamefont
  {Wilson}, \citenamefont {Wineland}, \citenamefont {Leibfried}, \citenamefont
  {Allcock},\ and\ \citenamefont {Slichter}}]{srinivas_2018}%
  \BibitemOpen
  \bibfield  {author} {\bibinfo {author} {\bibfnamefont {R.}~\bibnamefont
  {Srinivas}}, \bibinfo {author} {\bibfnamefont {S.~C.}\ \bibnamefont {Burd}},
  \bibinfo {author} {\bibfnamefont {R.~T.}\ \bibnamefont {Sutherland}},
  \bibinfo {author} {\bibfnamefont {A.~C.}\ \bibnamefont {Wilson}}, \bibinfo
  {author} {\bibfnamefont {D.~J.}\ \bibnamefont {Wineland}}, \bibinfo {author}
  {\bibfnamefont {D.}~\bibnamefont {Leibfried}}, \bibinfo {author}
  {\bibfnamefont {D.~T.~C.}\ \bibnamefont {Allcock}}, \ and\ \bibinfo {author}
  {\bibfnamefont {D.~H.}\ \bibnamefont {Slichter}},\ }\href@noop {} {\bibfield
  {journal} {\bibinfo  {journal} {Phys. Rev. Lett.}\ }\textbf {\bibinfo
  {volume} {122}},\ \bibinfo {pages} {163201} (\bibinfo {year}
  {2019})}\BibitemShut {NoStop}%
\bibitem [{\citenamefont {Sutherland}\ and\ \citenamefont
  {Erickson}(2024)}]{sutherland_2024}%
  \BibitemOpen
  \bibfield  {author} {\bibinfo {author} {\bibfnamefont {R.~T.}\ \bibnamefont
  {Sutherland}}\ and\ \bibinfo {author} {\bibfnamefont {S.~D.}\ \bibnamefont
  {Erickson}},\ }\href@noop {} {\bibfield  {journal} {\bibinfo  {journal}
  {Physical Review A}\ }\textbf {\bibinfo {volume} {109}},\ \bibinfo {pages}
  {022620} (\bibinfo {year} {2024})}\BibitemShut {NoStop}%
\bibitem [{\citenamefont {Edmonds}(1996)}]{edmonds_1996}%
  \BibitemOpen
  \bibfield  {author} {\bibinfo {author} {\bibfnamefont {A.~R.}\ \bibnamefont
  {Edmonds}},\ }\href@noop {} {\emph {\bibinfo {title} {Angular momentum in
  quantum mechanics}}}\ (\bibinfo  {publisher} {Princeton university press},\
  \bibinfo {year} {1996})\BibitemShut {NoStop}%
\bibitem [{\citenamefont {Langer}(2006)}]{langer_2006}%
  \BibitemOpen
  \bibfield  {author} {\bibinfo {author} {\bibfnamefont {C.~E.}\ \bibnamefont
  {Langer}},\ }\emph {\bibinfo {title} {High fidelity quantum information
  processing with trapped ions}},\ \href@noop {} {Ph.D. thesis},\ \bibinfo
  {school} {University of Colorado at Boulder} (\bibinfo {year}
  {2006})\BibitemShut {NoStop}%
\bibitem [{\citenamefont {Sutherland}\ \emph {et~al.}(2019)\citenamefont
  {Sutherland}, \citenamefont {Srinivas}, \citenamefont {Burd}, \citenamefont
  {Leibfried}, \citenamefont {Wilson}, \citenamefont {Wineland}, \citenamefont
  {Allcock}, \citenamefont {Slichter},\ and\ \citenamefont
  {Libby}}]{sutherland_2019}%
  \BibitemOpen
  \bibfield  {author} {\bibinfo {author} {\bibfnamefont {R.~T.}\ \bibnamefont
  {Sutherland}}, \bibinfo {author} {\bibfnamefont {R.}~\bibnamefont
  {Srinivas}}, \bibinfo {author} {\bibfnamefont {S.~C.}\ \bibnamefont {Burd}},
  \bibinfo {author} {\bibfnamefont {D.}~\bibnamefont {Leibfried}}, \bibinfo
  {author} {\bibfnamefont {A.~C.}\ \bibnamefont {Wilson}}, \bibinfo {author}
  {\bibfnamefont {D.~J.}\ \bibnamefont {Wineland}}, \bibinfo {author}
  {\bibfnamefont {D.~T.~C.}\ \bibnamefont {Allcock}}, \bibinfo {author}
  {\bibfnamefont {D.~H.}\ \bibnamefont {Slichter}}, \ and\ \bibinfo {author}
  {\bibfnamefont {S.~B.}\ \bibnamefont {Libby}},\ }\href@noop {} {\bibfield
  {journal} {\bibinfo  {journal} {New J. Phys.}\ }\textbf {\bibinfo {volume}
  {21}},\ \bibinfo {pages} {033033} (\bibinfo {year} {2019})}\BibitemShut
  {NoStop}%
\bibitem [{\citenamefont {Timoney}\ \emph {et~al.}(2011)\citenamefont
  {Timoney}, \citenamefont {Baumgart}, \citenamefont {Johanning}, \citenamefont
  {Var{\'o}n}, \citenamefont {Plenio}, \citenamefont {Retzker},\ and\
  \citenamefont {Wunderlich}}]{timoney_2011}%
  \BibitemOpen
  \bibfield  {author} {\bibinfo {author} {\bibfnamefont {N.}~\bibnamefont
  {Timoney}}, \bibinfo {author} {\bibfnamefont {I.}~\bibnamefont {Baumgart}},
  \bibinfo {author} {\bibfnamefont {M.}~\bibnamefont {Johanning}}, \bibinfo
  {author} {\bibfnamefont {A.~F.}\ \bibnamefont {Var{\'o}n}}, \bibinfo {author}
  {\bibfnamefont {M.~B.}\ \bibnamefont {Plenio}}, \bibinfo {author}
  {\bibfnamefont {A.}~\bibnamefont {Retzker}}, \ and\ \bibinfo {author}
  {\bibfnamefont {C.}~\bibnamefont {Wunderlich}},\ }\href@noop {} {\bibfield
  {journal} {\bibinfo  {journal} {Nature}\ }\textbf {\bibinfo {volume} {476}},\
  \bibinfo {pages} {185} (\bibinfo {year} {2011})}\BibitemShut {NoStop}%
\bibitem [{\citenamefont {S\'ark\'any}\ \emph {et~al.}(2014)\citenamefont
  {S\'ark\'any}, \citenamefont {Weiss}, \citenamefont {Hattermann},\ and\
  \citenamefont {Fort\'agh}}]{sark_2014}%
  \BibitemOpen
  \bibfield  {author} {\bibinfo {author} {\bibfnamefont {L.}~\bibnamefont
  {S\'ark\'any}}, \bibinfo {author} {\bibfnamefont {P.}~\bibnamefont {Weiss}},
  \bibinfo {author} {\bibfnamefont {H.}~\bibnamefont {Hattermann}}, \ and\
  \bibinfo {author} {\bibfnamefont {J.}~\bibnamefont {Fort\'agh}},\ }\href
  {\doibase 10.1103/PhysRevA.90.053416} {\bibfield  {journal} {\bibinfo
  {journal} {Phys. Rev. A}\ }\textbf {\bibinfo {volume} {90}},\ \bibinfo
  {pages} {053416} (\bibinfo {year} {2014})}\BibitemShut {NoStop}%
\bibitem [{\citenamefont {Srinivas}\ \emph {et~al.}(2023)\citenamefont
  {Srinivas}, \citenamefont {L{\"o}schnauer}, \citenamefont {Malinowski},
  \citenamefont {Hughes}, \citenamefont {Nourshargh}, \citenamefont
  {Negnevitsky}, \citenamefont {Allcock}, \citenamefont {King}, \citenamefont
  {Matthiesen}, \citenamefont {Harty} \emph {et~al.}}]{srinivas_2023}%
  \BibitemOpen
  \bibfield  {author} {\bibinfo {author} {\bibfnamefont {R.}~\bibnamefont
  {Srinivas}}, \bibinfo {author} {\bibfnamefont {C.}~\bibnamefont
  {L{\"o}schnauer}}, \bibinfo {author} {\bibfnamefont {M.}~\bibnamefont
  {Malinowski}}, \bibinfo {author} {\bibfnamefont {A.}~\bibnamefont {Hughes}},
  \bibinfo {author} {\bibfnamefont {R.}~\bibnamefont {Nourshargh}}, \bibinfo
  {author} {\bibfnamefont {V.}~\bibnamefont {Negnevitsky}}, \bibinfo {author}
  {\bibfnamefont {D.}~\bibnamefont {Allcock}}, \bibinfo {author} {\bibfnamefont
  {S.}~\bibnamefont {King}}, \bibinfo {author} {\bibfnamefont {C.}~\bibnamefont
  {Matthiesen}}, \bibinfo {author} {\bibfnamefont {T.}~\bibnamefont {Harty}},
  \emph {et~al.},\ }\href@noop {} {\bibfield  {journal} {\bibinfo  {journal}
  {Physical Review Letters}\ }\textbf {\bibinfo {volume} {131}},\ \bibinfo
  {pages} {020601} (\bibinfo {year} {2023})}\BibitemShut {NoStop}%
\end{thebibliography}%

\end{document}